\documentclass[9pt,twocolumn,twoside]{osajnl}

\journal{ol} 

\usepackage{siunitx}

\setboolean{shortarticle}{true}

\title{23 mJ high-power fiber CPA system using electro-optically controlled divided-pulse amplification}

\author[1,*]{Henning Stark}
\author[1]{Joachim Buldt}
\author[1]{Michael M\"uller}
\author[1,2]{Arno Klenke}
\author[1,2,3]{Andreas T\"unnermann}
\author[1,2,3]{Jens Limpert}

\affil[1]{Institute of Applied Physics, Abbe Center of Photonics, Friedrich-Schiller-University Jena, Albert-Einstein-Straße 15, 07745 Jena, Germany}
\affil[2]{Helmholtz-Institute Jena, Fröbelstieg 3, 07743 Jena, Germany}
\affil[3]{Fraunhofer Institute for Applied Optics and Precision Engineering, Albert-Einstein-Str. 7, 07745 Jena, Germany}

\affil[*]{Corresponding author: lars.henning.stark@uni-jena.de}

\dates{}

\doi{\url{https://doi.org/10.1364/OL.44.005529}}

\begin{abstract}
The pulse-energy scaling technique electro-optically controlled divided-pulse amplification is implemented in a high-power ultrafast fiber laser system based on coherent beam combination. A fiber-integrated front end and a multi-pass cell based back end allow for a small footprint and a modular implementation. Bursts of 8 pulses are amplified parallel in up to 12 ytterbium-doped large-pitch fiber amplifiers. Subsequent spatio-temporal coherent combination of the 96 total amplified pulse replicas to a single pulse results in a pulse energy of 23 mJ at an average power of 674 W, compressible to a pulse duration of 235 fs. To the best of our knowledge, this is the highest pulse energy ever accomplished with a fiber CPA system.
\end{abstract}

\setboolean{displaycopyright}{true}

\begin{document}

\maketitle

\noindent High-power ultrafast laser systems are versatile tools for a variety of applications, such as THz or XUV generation for spectroscopy and imaging, demanding for ever higher average power, peak power and pulse energy. On the one hand, to provide such challenging output parameters, fiber lasers are widely appreciated for their high average power capability, a rather simple setup and an excellent beam quality. On the other hand, transversal mode-instabilities and nonlinear effects are known limitations \cite{Jauregui2013}. Although widely applied power-scaling approaches such as large-mode-area fiber schemes (LMA) \cite{Stutzki2014} and chirped-pulse amplification (CPA) \cite{Strickland1985} allowed for a significant increase of the achievable performance, their potential is already well exploited. Thus, further scaling requires new technologies. Coherent beam combination (CBC), for instance, is commonly seen as an artificial mode-area scaling and has shown remarkable results in the recent past \cite{Mueller2018}. As a time-domain counterpart, techniques have been developed to aim for an additional increase of the already CPA-stretched pulse duration during amplification. These pulse duration enhancing approaches share a similar underlying principle: Multiple consecutive pulses are amplified instead of a single one, potentially increasing the energy extraction and, after temporal combination into a single pulse, allowing to generate formerly unmatched pulse energies. 
 
The most famous techniques following this concept are coherent pulse stacking (CPS) \cite{Zhou2015}, cavity enhancement \cite{Breitkopf2014} and divided-pulse amplification (DPA) \cite{Zhou2007}. Previously, DPA has shown its potential especially in an experiment comprising 4 temporal pulse replicas and CBC of 8 amplifier channels, generating pulse energies of $12\,\mathrm{mJ}$ at an average power of $700\,\mathrm{W}$ \cite{Kienel2016}. Additionally, the experiment revealed that 8 or even 16 pulse replicas would be reasonable for efficient energy extraction for the given pulse duration and fiber type. However, conventional DPA cannot support more than 4 pulse replicas efficiently due to a limited access to phase and amplitude of the individual pulses which are required for compensation of amplifier saturation and nonlinearity. A new approach named electro-optically controlled divided-pulse amplification (EDPA) offers control over amplitude and phase of each individual pulse of the burst in a compact fiber-integrated frontend and has already shown promising results in a first proof-of-principle experiment \cite{Stark2017}.

In this contribution, we present the results of the first implementation of EDPA in an existing high-power ultrafast fiber laser system based on CBC, finally comprising a spatio-temporal pulse addition of 8-pulse bursts from up to 12 parallel amplifiers.

\begin{figure}[tb]
	\begin{center}
		\includegraphics[width=\linewidth*40/45]{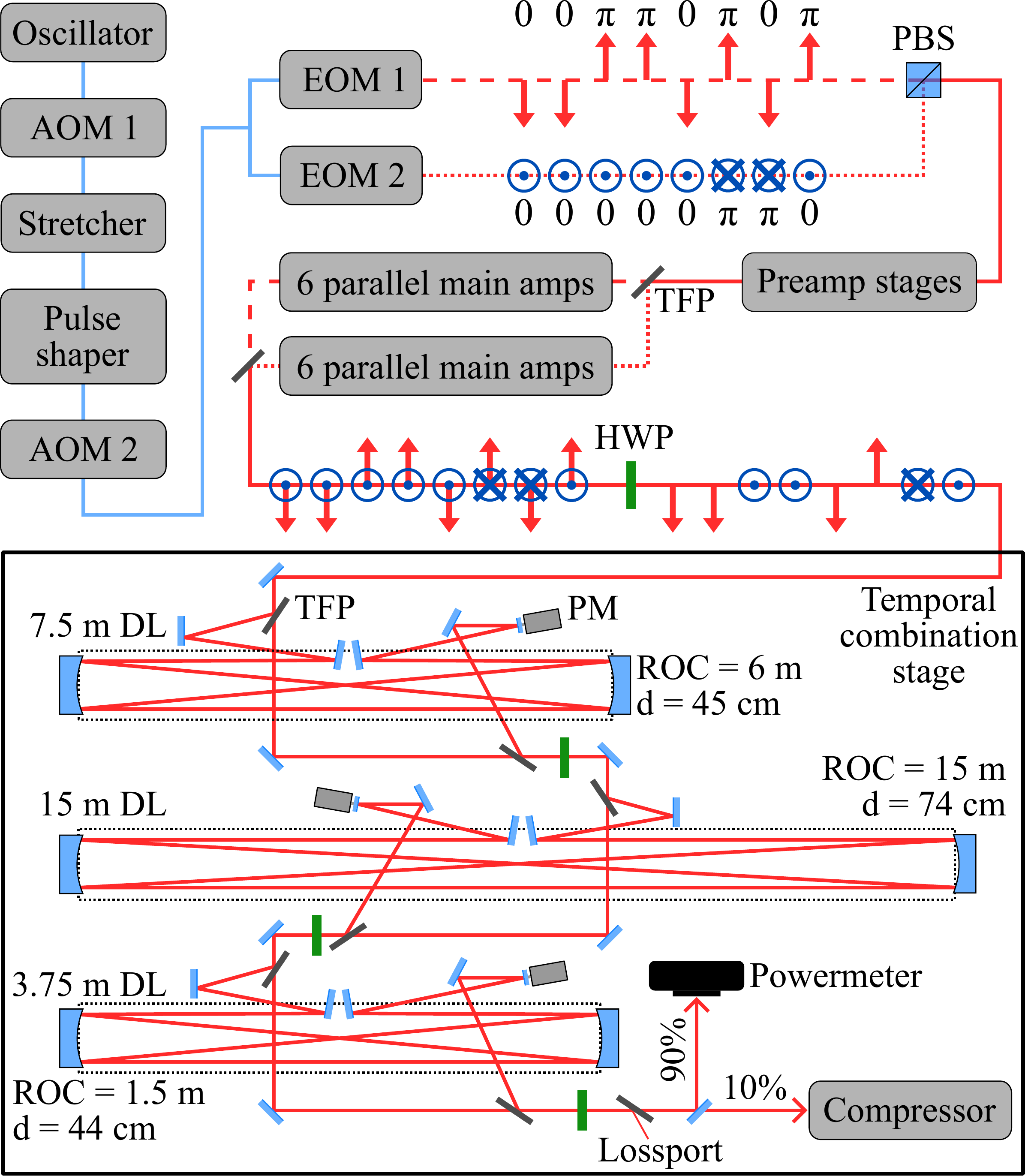}
		\caption{Simplified setup for spatio-temporal combination of 12 parallel main amplifiers and 8 temporal pulse replicas. AOM: acousto-optic modulator, EOM: electro-optic modulator, PBS: polarizing beam splitter, TFP: thin-film polarizer, HWP: half-wave plate, PM: piezo-mounted mirror, DL: delay line, ROC: radius of curvature of MPC mirrors, d: MPC mirror distance.}
		\label{fig:setup}
	\end{center}
\end{figure}

The experimental setup is depicted in Fig.~\ref{fig:setup} and is based on the ultrafast high-power CBC system presented in \cite{Mueller2016_b, MuellerPW2018}. In this experiment, a passively mode-locked Yb:KYW oscillator is used as seed source, emitting femtosecond pulses at a repetition rate of $\SI{80}{MHz}$. A fiber-coupled acousto-optic modulator (AOM~1) picks bursts of 8 consecutive pulses at a rate of $f_\mathrm{burst}=\SI{1}{MHz}$, referred to as burst repetition rate. An Oeffner-type stretcher with a hardcut of $\SI{22}{nm}$ and a center wavelength of $\SI{1030}{nm}$ stretches the pulse duration to $\SI{1.9}{ns}$ FWHM. A set of 4 commercially available spectral transmission notch filters counteracts gain narrowing and a spatial light modulator (SLM) in conjunction with the multiphoton intrapulse interference phase scan technique \cite{Lozovoy2004} compensates for accumulated nonlinear phase. This is followed by a second fiber-coupled AOM (AOM~2), which is controlled by an arbitrary waveform generator (AWG). Its dual purpose is the reduction of the burst repetition rate to a final value and amplitude pre-shaping within the bursts to compensate for gain saturation in the main amplifiers. The resulting signal is divided equally into two identical fiber channels, each containing an electro-optic phase-modulator (EOM) with $\SI{5}{GHz}$ bandwidth. A 2-channel AWG ($\SI{1.2}{GS/s}$) uses the EOMs to imprint distinctive phase patterns on the bursts. By subsequent coherent polarization combination of both signals a burst with any desired pattern of orthogonally polarized pulses can be created. In particular, this ultimately allows to generate the polarization pattern as required for the temporal pulse combination \cite{Stark2017}.

The modifications necessary in the front end are minimal, involving basically only the addition of the two EOM channels. The output signals of both EOM channels (dotted and dashed lines in Fig.~\ref{fig:setup}) are spatially superimposed in free space with orthogonal polarizations and a slight temporal delay, using a polarizing beam splitter (PBS) and a path length difference of about $\SI{17}{cm}$. Since this delay significantly exceeds the coherence length of the pulses, both signals can be sent simultaneously through the following preamplifier system without noticeably affecting each other. The preamplifier system consists of two subsequent ytterbium-doped large-pitch fiber (LPF) amplifiers generating an average power of $\SI{11}{W}$. Then, a thin-film polarizer (TFP) again separates both polarizations, i.e. the amplified signals of the individual EOM-channels, indicated as the dotted and the dashed lines again. Each signal is used to seed one group of up to 6 parallel ytterbium-doped LPF main amplifiers each. Then, the 6 amplified output beams within the respective amplifier groups are coherently combined by means of thin-film polarizers (TFP). For the necessary active phase stabilization the H\"ansch-Couillaud polarization detection method \cite{Hansch1980} in conjunction with piezo-mounted mirrors on the seed side of the main amplifiers are applied. Finally, the two amplifier-groups themselves are interferometrically combined, generating an 8-pulse burst with the pattern of alternating orthogonal polarizations as required for the following temporal combination. 

The temporal combination stage is based on fused-silica TFPs as spatial splitting and combining elements and three optical delay lines with lengths of $\SI{3.75}{m}$, $\SI{7.5}{m}$ and $\SI{15}{m}$, which are matched to the $80\,\mathrm{MHz}$ repetition rate of the laser oscillator. Each delay line contains a Herriott-type multi-pass cell (MPC) \cite{Herriott1965}, allowing for a small overall footprint of about $\SI{0.5}{m^2}$ including all delay lines. To limit the fluences on the mirrors, the delay lines were ordered by the minimum beam diameters occurring on the MPC mirrors (Fig. \ref{fig:setup}). Finally, the combined signal passes a last TFP, which removes most of the uncombined signal from the beam. An outcoupling mirror then reflects 90\% of the combined power onto a powermeter and transmits 10\% into the compressor. This sampling is required as the total pulse energies aimed at are incompatible with recompression and subsequent propagation in the ambient air compressor. The reason for this is expected to be ionization and thermal load in the air induced by the high peak and average power of the compressed pulses, spoiling the laser beam critically. A new compressor, capable of operating at this power level, is currently under construction.

LOCSET \cite{Shay2006} is used as a phase stabilization scheme for the last spatial combination step and all of the following temporal combination steps. The required phase dithers have frequencies between $\SI{3}{kHz}$ and $\SI{9}{kHz}$ and are applied by piezo-mounted mirrors on the seed side between both 6-amplifier groups and in each delay line. A single photodetector after the last TFP detects the corresponding phase mismatches. Additionally, a timing gate is implemented into the error signal detection to ensure correct stabilization, since phase locking in DPA and CPS is known to exhibit multiple stable states \cite{Mueller2016_a}.

\begin{figure}[tb]
	\begin{center}
		\includegraphics[width=\linewidth*8/11]{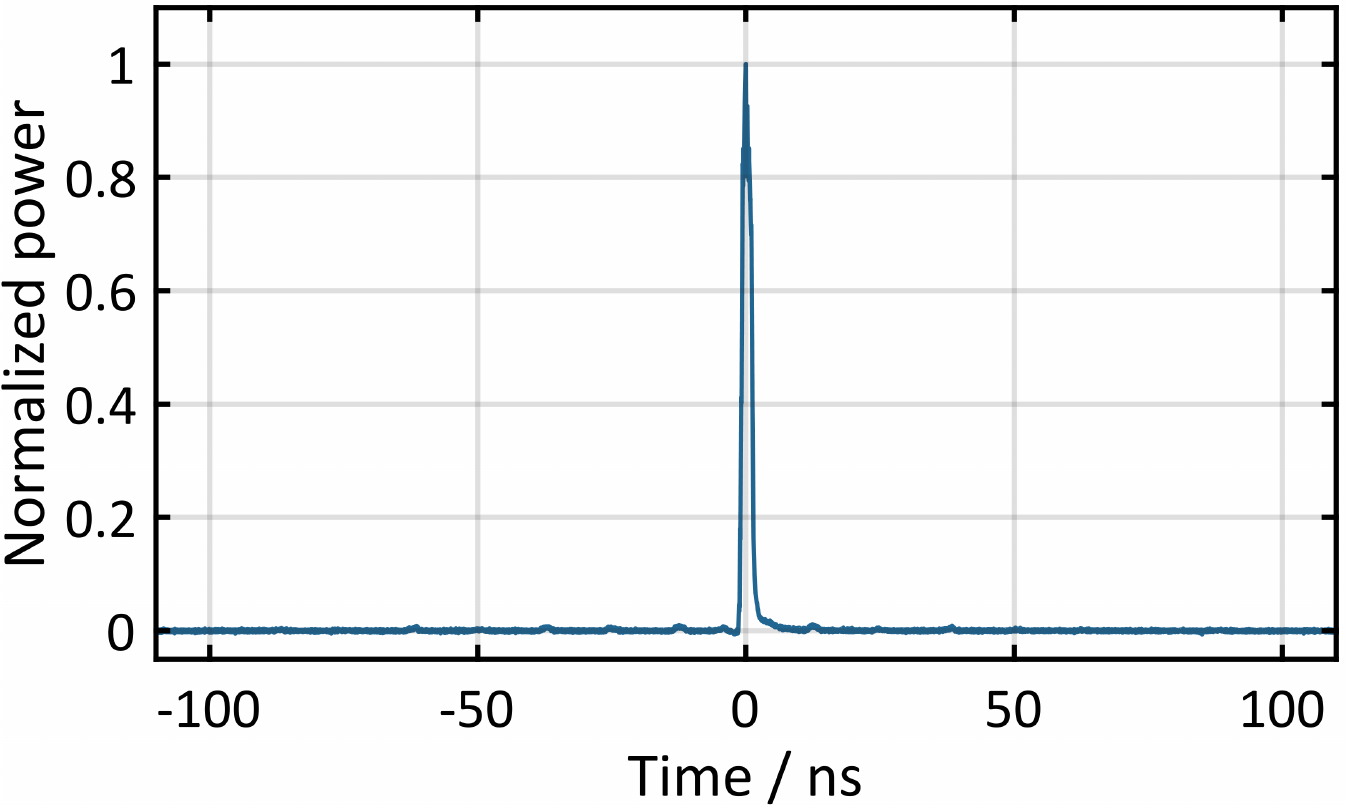}
		\caption{Photo diode trace of the combined signal from two amplifiers at a burst repetition rate of $f_\mathrm{burst}=\SI{500}{kHz}$.}
		\label{fig:500kHz_comb}
	\end{center}
\end{figure}

For an initial investigation and maximization of the attainable combining performance, the number of parameters which have to be optimized is desired to be as small as possible. Thus, the burst repetition rate is set to a high value of $\SI{500}{kHz}$, minimizing the impact of gain saturation and nonlinear phase accumulation, rendering amplitude pre-shaping unnecessary. For this, the output signals of two main amplifiers, one of each amplifier group, are temporally and spatially combined. The resulting signal is depicted in Fig.~\ref{fig:500kHz_comb} and it has an average power of $\SI{140}{W}$. The combining efficiency is $\eta_\mathrm{comb}=83\%$, denoting the ratio between the average power of the linearly polarized, temporally combined signal after the very last TFP and the average power available for temporal combining. The temporal efficiency is $\eta_\mathrm{temp} = 94\%$, meaning that 94\% of the combined power is contributed by the main pulse and 6\% by pre- and post-pulses. These side features are the result of an imperfect temporal combination and can be caused by, for instance, remaining amplitude and phase mismatches within the amplified bursts or by misaligned delay lines. The stacking efficiency in Eq.~\ref{eq:eta_stack} combines both figures of merit and is defined here as the ratio of the energy in the main pulse divided by the total energy available for temporal combination, which is the energy per burst in front of the temporal combining stage. Consequently, the stacking efficiency amounts to 
\begin{equation}
\label{eq:eta_stack}
\eta_\mathrm{stack} = \frac{E_\mathrm{main}}{\sum_{i} E_i} = \eta_\mathrm{comb}\eta_\mathrm{temp} =  78\%\; 
\end{equation}
with $i$ being the index of the individual temporal pulse replicas before temporal combining. 

Another figure of merit used to measure the temporal combination quality is the pre-pulse contrast $C$. It is defined as the ratio of energy in the main pulse $E_\mathrm{main}$ to the energy in the most intense pre-pulse $E_\mathrm{pre}$. It is calculated by evaluation of the photodiode trace of the combined signal, resulting in 
\begin{equation}
\label{eq:contrast}
C=\frac{E_\mathrm{main}}{E_\mathrm{pre}}=\SI{21}{dB}\; .
\end{equation}

Next, the burst repetition rate is reduced to $\SI{25}{kHz}$, consequently increasing the achievable energy per burst. However, the impact of gain saturation also becomes significantly stronger, causing a steep slope of the pulse energies within the burst. Without amplitude pre-shaping, this amplitude mismatch of the pulses and the accompanying mismatch of the accumulated nonlinear phase substantially reduce the temporal combination performance. Furthermore, the first pulse of the burst contains by far the highest energy. Thus, since all individual pulses must stay below the single pulse damage threshold of the amplifier fibers, the first pulse limits the energy extractable by the whole burst to a low level. To counteract these performance impairing effects, amplitude pre-shaping is applied. For this, AOM~2 in the fiber front end is used, providing control over the amplitudes of all individual pulse replicas prior to main amplification. 

For a simple matching of the nonlinear phase, which has been shown to be the most critical parameter for efficient combining \cite{Kienel2016}, another advantage of EDPA is exploited: The pulse polarization pattern can be freely selected by adapting the EOM phase patterns and, for instance, a polarization pattern consisting of 1 p-polarized pulse and 7 s-polarized pulses can be directly generated. Thus, blocking of the first delay line allows to extinguish these 7 pulses, while only a single pulse continues. By turning the following HWPs the remaining pulse is transmitted by all TFPs and can be measured in the end, for example by an auto-correlator. This approach is applied to single out each individual pulse of the amplified burst one by one at the target pulse energy. 

\begin{figure}[tb]
	\begin{center}
		\includegraphics[width=\linewidth*11/24]{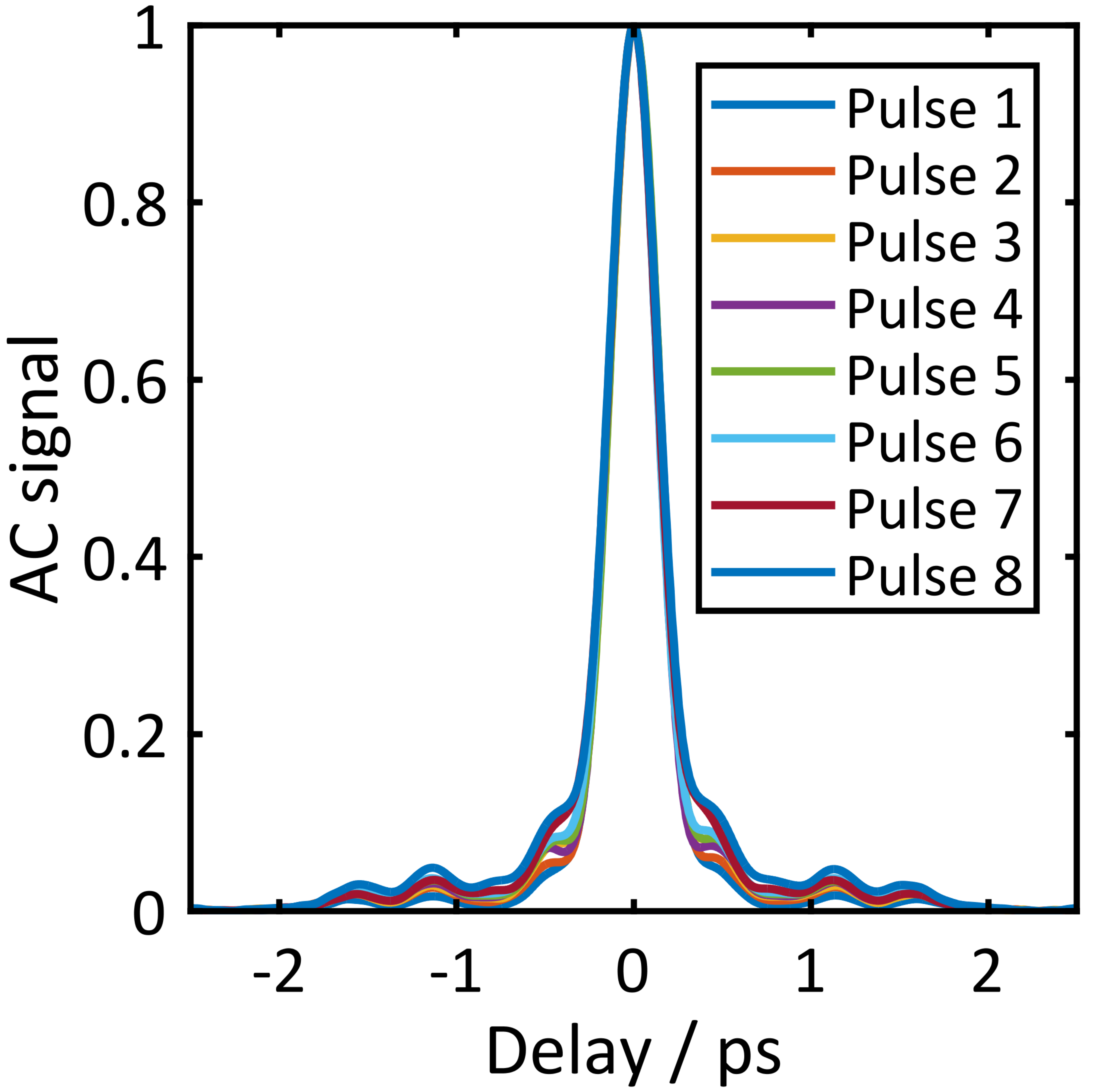}
		\caption{Matched auto-correlation traces of the individual pulses of the amplified burst after amplitude pre-shaping.}
		\label{fig:AC_matching}
	\end{center}
\end{figure}

\begin{figure}[tb]
	\begin{center}
		\includegraphics[width=\linewidth*8/11]{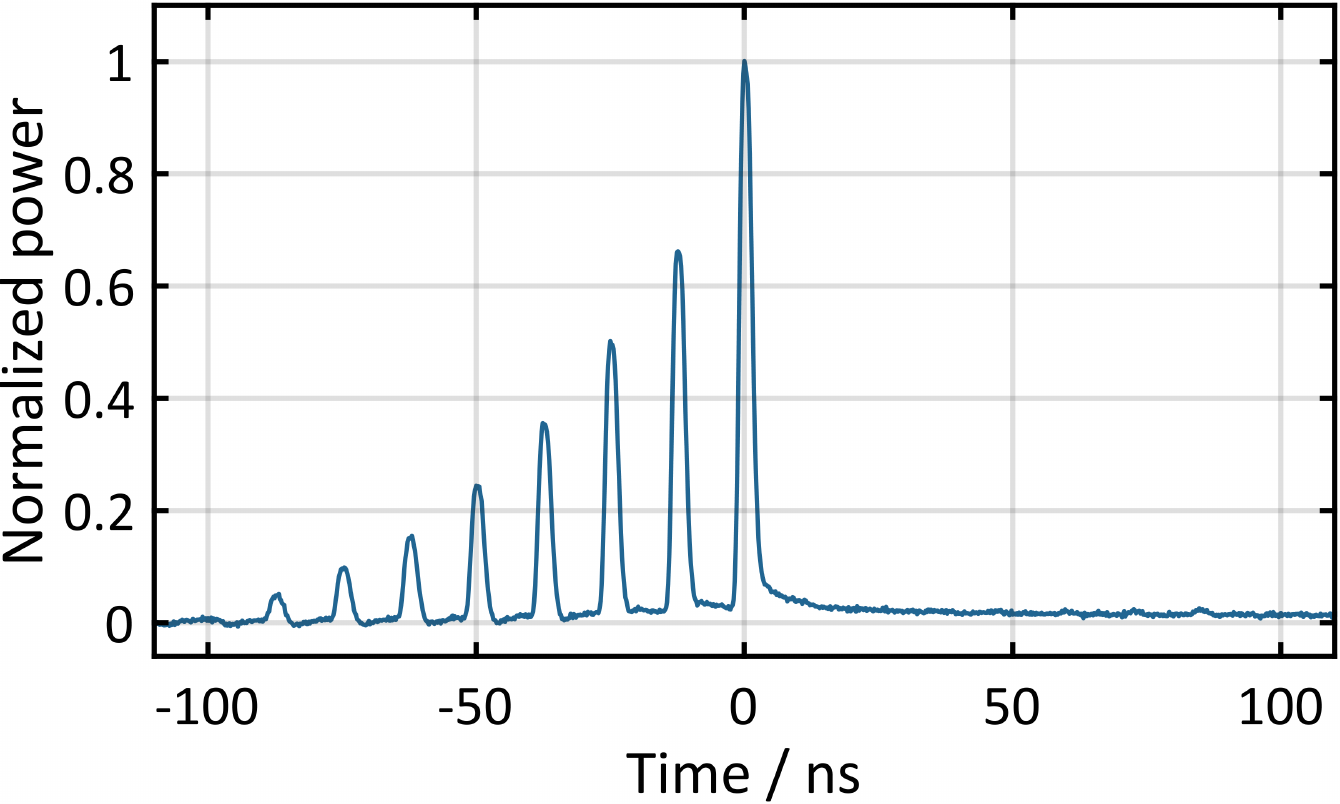}
		\caption{Seed signal burst after amplitude pre-shaping at $f_\mathrm{burst}=\SI{25}{kHz}$ measured with a slower photo diode.}
		\label{fig:seed_pre-shaping}
	\end{center}
\end{figure}

\begin{figure}[tb!]
	\begin{center}
		\includegraphics[width=\linewidth*8/11]{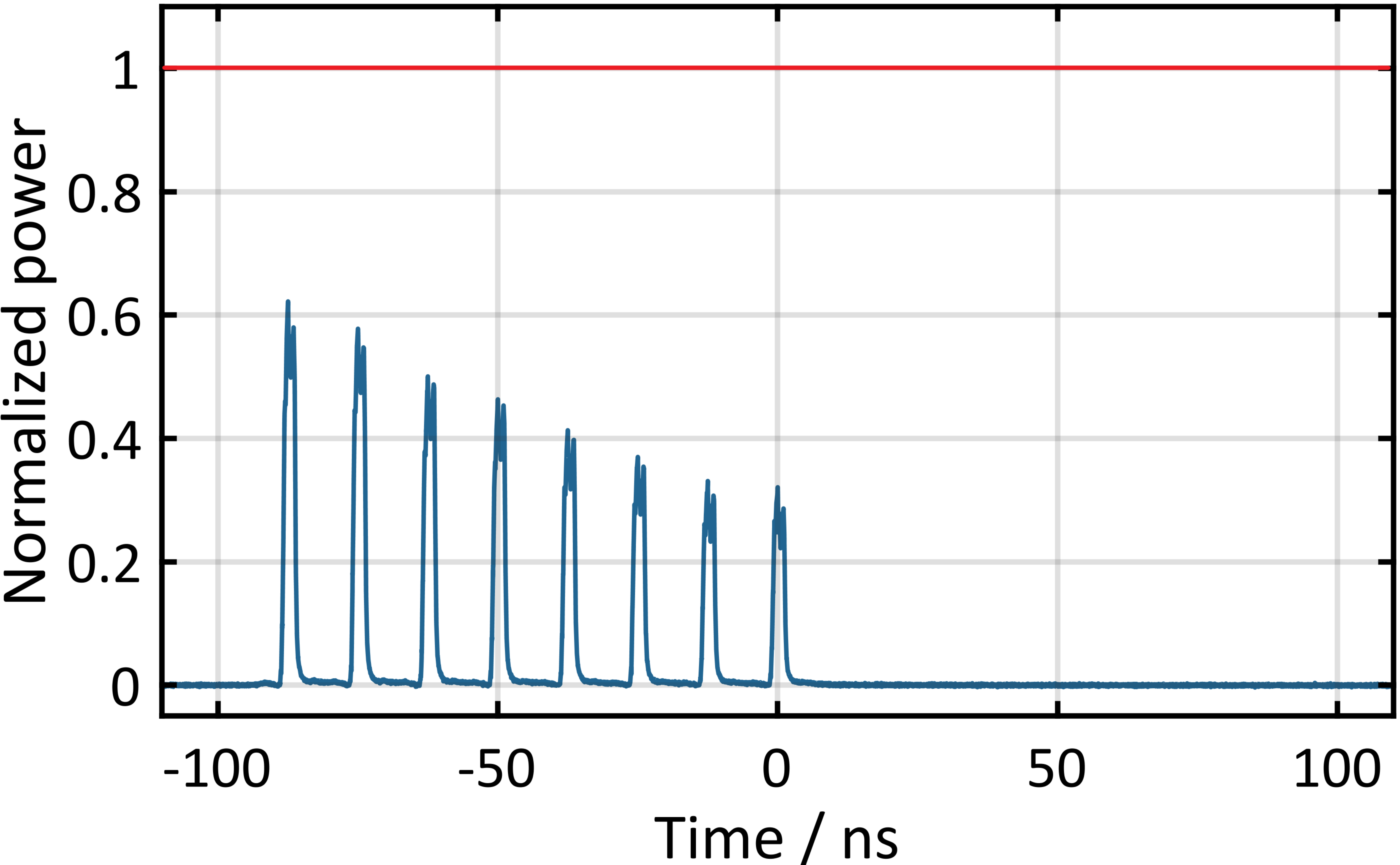}
		\caption{Burst after main amplification at $f_\mathrm{burst}=\SI{25}{kHz}$ normalized to the corresponding estimated single pulse damage threshold of an amplifier at about $\SI{1}{mJ}$, indicated by a red line.}
		\label{fig:amplified_burst}
	\end{center}
\end{figure}

The compressed pulse duration of the first pulse is minimized to $\SI{235}{fs}$, assuming a Gaussian pulse shape, by the SLM in the front end. Next, amplitude pre-shaping is used to match the individual auto-correlation traces of the following pulse replicas, as depicted in Fig.~\ref{fig:AC_matching}, implying similar accumulated nonlinear phases. The resulting seed pulse train measured prior to the LPF pre-amplifiers is shown in Fig.~\ref{fig:seed_pre-shaping}. The corresponding amplified burst contains about $\SI{3.6}{mJ}$ per channel and is shown in Fig.~\ref{fig:amplified_burst}, measured by a photodiode with $\SI{12}{ps}$ rise time and an $\SI{80}{GS/s}$ oscilloscope. The substantial difference between the slopes from the amplified burst and the seed burst indicates a strong gain saturation and, thus, a significant extent of energy extraction.

The spatial and temporal combination of two channels results in an average combined power of $\SI{143}{W}$ with a combining efficiency of 80\%. The pre- and post-pulses are slightly more pronounced, leaving a temporal contrast of about $\SI{19}{dB}$ and a temporal efficiency of more than 90\%. This marginal deterioration of the combining performance can be assigned to remaining pulse-to-pulse mismatches of phase and amplitude as well as spectral differences at this high level of gain saturation. 

\begin{figure}[tb]
	\begin{center}
		\includegraphics[width=\linewidth*8/11]{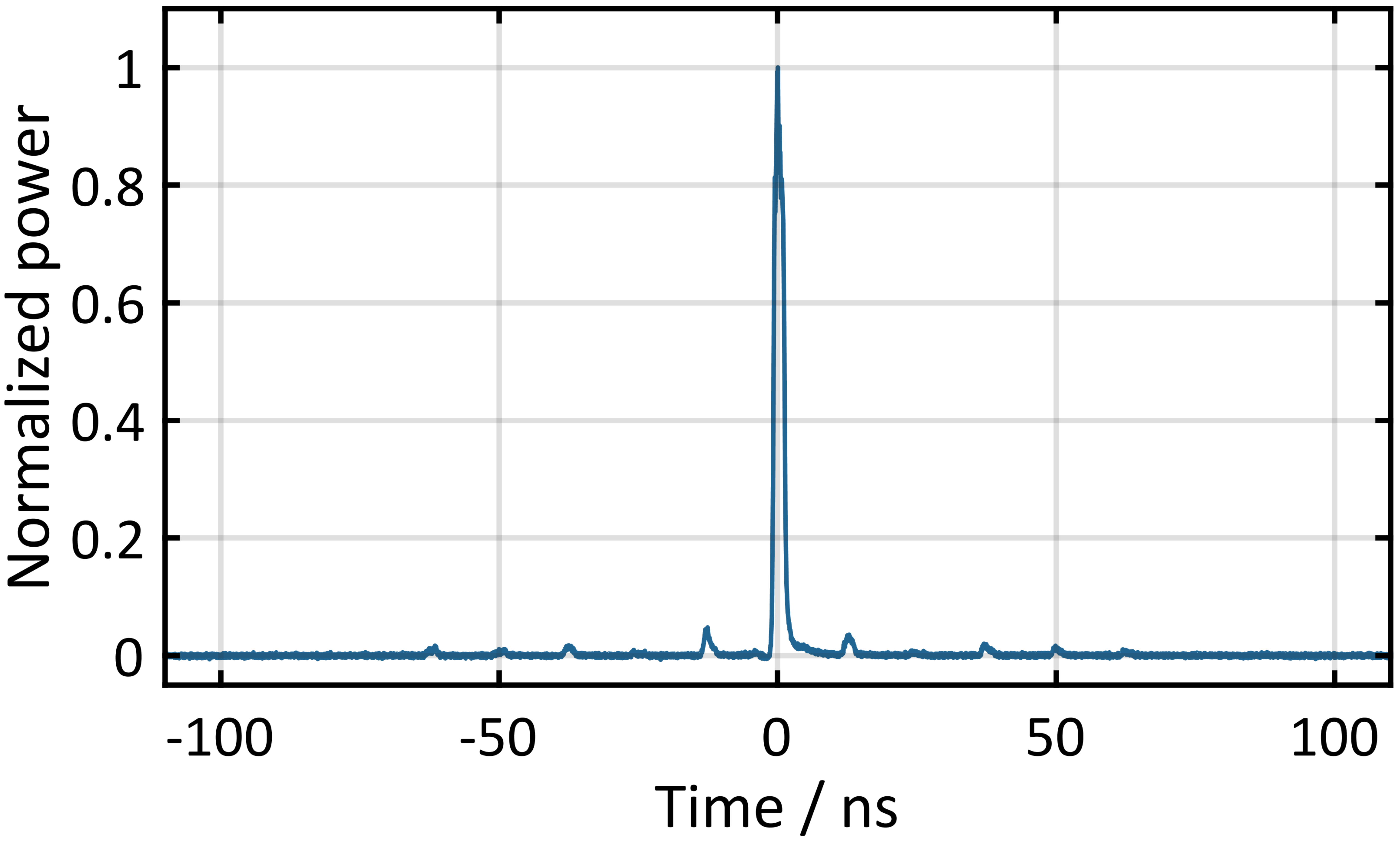}
		\caption{Photo diode trace of the spatio-temporally combined signal from 12 main amplifiers with $\SI{23}{mJ}$ in the main feature.}
		\label{fig:PD_25kHz_12CH}
	\end{center}
\end{figure}

\begin{figure}[tb]
	\begin{center}
		\includegraphics[width=\linewidth*8/11]{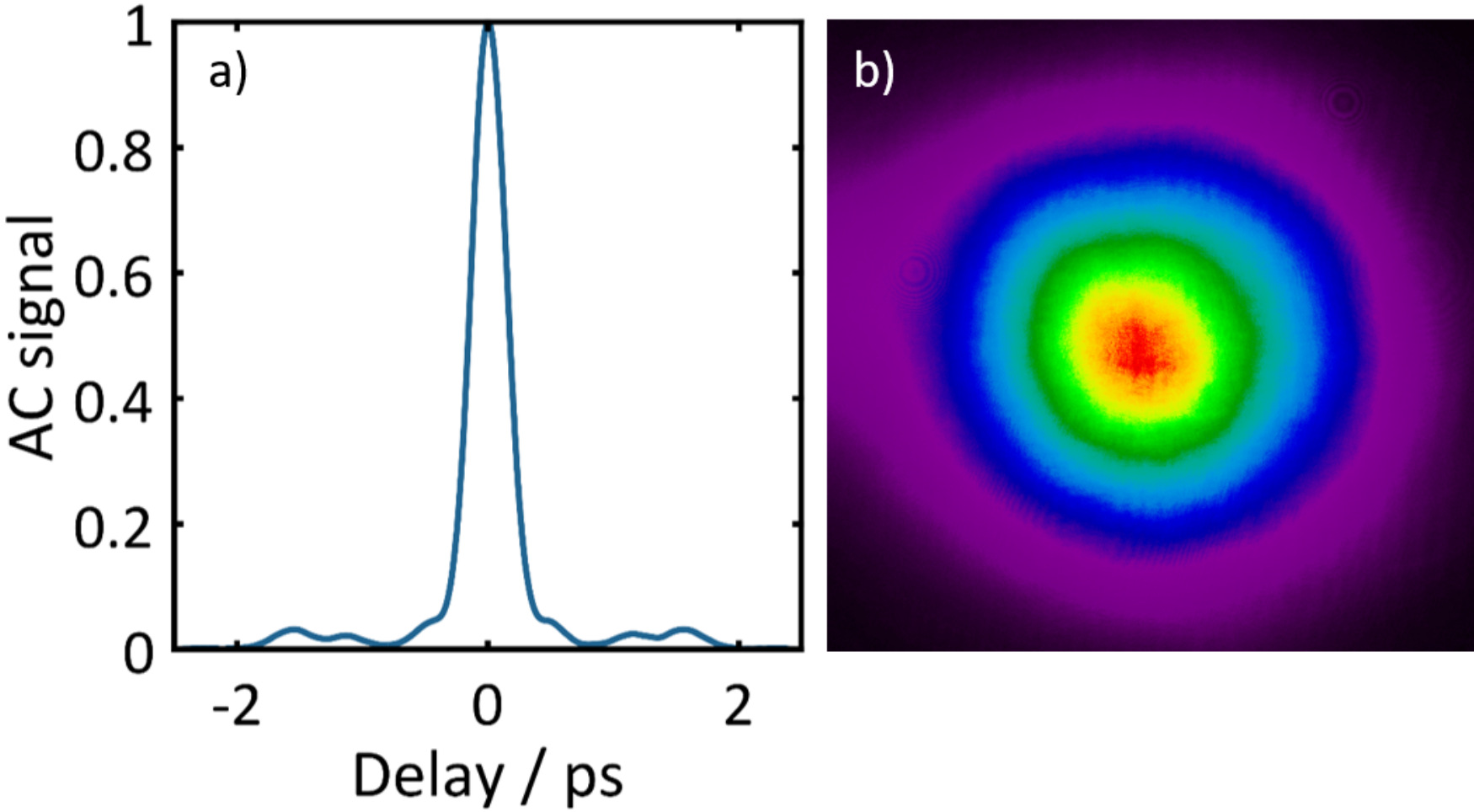}
		\caption{a) Auto-correlation trace of the combined pulse sample after compression. b) Intensity profile of the combined beam.}
		\label{fig:AC_IP}
	\end{center}
\end{figure}

The next experiment involves 12 main amplifier channels and, consequently, a spatio-temporal distribution into 96 individual pulses during amplification. The fully combined signal reaches $\SI{674}{W}$ of average power at a combining efficiency of 71\%. The photo-diode trace of the uncompressed signal is depicted in Fig.~\ref{fig:PD_25kHz_12CH}. The pulse contains a total energy of $\SI{27}{mJ}$, including pre- and post-pulses. The energy in the main feature is $\SI{23}{mJ}$ and the temporal efficiency exceeds 85\%. The temporal contrast is $\SI{15}{dB}$. The auto-correlation trace of the compressed sample is shown in Fig.~\ref{fig:AC_IP} a) and has a pulse duration of $\SI{235}{fs}$, which equals the pulse durations of the individual pulses. Furthermore, also the Gaussian intensity profile of the beam is maintained after the combination, as depicted in Fig.~\ref{fig:AC_IP} b). The achieved output parameters are summarized in Table \ref{tab:1}.

\begin{table}[h!]
	\centering
	\begin{tabular}{ccccccc}
		\hline 
		$f_\mathrm{burst}$ & ch. & $P_\mathrm{comb}$ & $\eta_\mathrm{sys}$ & $\eta_\mathrm{comb}$ & $C$ & $E_\mathrm{main}$ \\ 
		\hline 
		$\SI{500}{kHz}$ & 2 & $\SI{140}{W}$ & 78\% & 83\% & 21\,dB & $\SI{0.26}{mJ}$ \\ 
		$\SI{25}{kHz}$ & 2 & $\SI{143}{W}$ & 72\% & 80\% & 19\,dB & $\SI{5.2}{mJ}$ \\ 
		$\SI{25}{kHz}$ & 12 & $\SI{674}{W}$ & 56\% & 71\% & 15\,dB & $\SI{23}{mJ}$ \\ 
		\hline 
	\end{tabular} 

	\caption{Summarized results. The system efficiency $\eta_\mathrm{sys}$ includes the stacking efficiency and the spatial combining efficiency within both amplifier groups (for 12 channels). It is the energy in the finally combined main pulse divided by the sum of energies emitted by the single amplifiers. ch: channel count.} 
	\label{tab:1}
\end{table}

Efficiencies and temporal contrast are slightly reduced compared to the previous experiment. This is attributed to the onset of thermal lensing in the TFPs. While the un-delayed pulses in each temporal combination step are transmitted by the TFPs, the delayed beams are reflected by the very same ones. Thus, both beams see a very weak but opposite wavefront change when noticeable heating of the TFPs arises. This effect accumulates over the 6 involved TFPs. It is a mere technical issue, which will most probably be solved by replacing the TFPs with new high-power capable ones. 

In summary, we presented the first implementation of the pulse-energy scaling technique electro-optically controlled divided-pulse amplification in a high-power ultrafast fiber laser system based on CBC. At a burst repetition rate of $\SI{25}{kHz}$ about $\SI{3.6}{mJ}$ are extracted per amplifier fiber by bursts of 8 pulses, which is nearly 4 times the estimated single pulse damage threshold of the fiber. Using 2 main amplifiers, a combination efficiency of 80\% and a high temporal contrast of $\SI{19}{dB}$ are verified. The pulse duration of the combined and compressed pulse is $\SI{235}{fs}$, which equals the pulse durations of the individual pulse replicas within a burst. Using 12 parallel main amplifiers, a combined power of $\SI{674}{W}$ and an energy of $\SI{23}{mJ}$ in the main pulse are achieved. This is, to the best of our knowledge, the highest pulse energy ever achieved with a fiber CPA system to date. A sample of the combined pulse is compressed to $\SI{235}{fs}$ pulse duration which, when assuming a future compression of the full energy in a high-efficiency dielectric grating compressor with 90\% transmission, would correspond to about $\SI{80}{GW}$ peak power. The combining efficiency and the temporal contrast are slightly reduced to 71\% and $\SI{15}{dB}$, respectively, most likely due to onsetting thermal lensing in the temporal combining TFPs. 

In future iterations, the TFPs will be replaced and the combining setup will be optimized for high powers. Extended detection and enhanced analysis of the individual combination steps will allow for a better adjustment of the setup, thus further increasing the overall efficiency and temporal contrast. Additionally, a new large-aperture compressor with inert gas atmosphere will enable to compress the entire output signal. Ultimately, we are confident that EDPA will allow to generate ultrashort pulses with $\SI{100}{mJ}$-level pulse energy at multi-kW average power with existing fiber-based amplification setups in the near future.

\section*{Funding}
Th\"uringer Aufbaubank and European Regional Development Fund (ERDF) (2015FE9158); European Research Council (617173, 670557).

\bibliography{references}

\ifthenelse{\equal{\journalref}{aop}}{%
\section*{Author Biographies}
\begingroup
\setlength\intextsep{0pt}
\begin{minipage}[t][6.3cm][t]{1.0\textwidth} 
  \begin{wrapfigure}{L}{0.25\textwidth}
    \includegraphics[width=0.25\textwidth]{john_smith.eps}
  \end{wrapfigure}
  \noindent
  {\bfseries John Smith} received his BSc (Mathematics) in 2000 from The University of Maryland. His research interests include lasers and optics.
\end{minipage}
\begin{minipage}{1.0\textwidth}
  \begin{wrapfigure}{L}{0.25\textwidth}
    \includegraphics[width=0.25\textwidth]{alice_smith.eps}
  \end{wrapfigure}
  \noindent
  {\bfseries Alice Smith} also received her BSc (Mathematics) in 2000 from The University of Maryland. Her research interests also include lasers and optics.
\end{minipage}
\endgroup
}{}

\end{document}